\documentclass[sigconf,natbib]{acmart}
\AtBeginDocument{%
  \providecommand\BibTeX{{%
    \normalfont B\kern-0.5em{\scshape i\kern-0.25em b}\kern-0.8em\TeX}}}

\acmYear{2023}
\acmDOI{XXXXXXX.XXXXXXX}

\usepackage{balance}

\begin{document}

\title{Effectiveness and Efficiency Trade-off in Selective Query Processing}

\author{Josiane Mothe}
 \orcid{0000-0001-9273-2193}
 \affiliation{IRIT UMR5505 CNRS, INSPE, Univ. de Toulouse
  \institution{ }
  \streetaddress{}
  \city{Toulouse} 
  \country{France}
\postcode{F-31062}
}
\email{josiane.mothe@irit.fr}

\author{Md Zia Ullah}
\orcid{0000-0002-4022-7344}
 \affiliation{SCEBE, Edinburgh Napier University
 \institution{ }
 \streetaddress{}
 \city{Edinburgh}
 \country{United Kingdom}
 \postcode{EH10 5DT}
}
\email{m.ullah@napier.ac.uk}

\renewcommand{\shortauthors}{Trovato and Tobin, et al.}

\begin{abstract}
 Query processing in search engines can be optimized for use for all queries. For this, system component parameters such as the weighting function or the automatic query expansion model can be optimized or learned from past queries. However, it may be more interesting to optimize the processing thread on a query-by-query basis by adjusting the component parameters; this is what selective query processing does. Selective query processing uses one of the candidate processing threads chosen at query time. The choice is based on query features. In this paper, we examine selective query processing in different settings, both in terms of effectiveness and efficiency; this includes selective query expansion and other forms of selective query processing (e.g., when the term weighting function varies or when the expansion model varies). We found that the best trade-off between effectiveness and efficiency is obtained when using the best trained processing thread and its query expansion counter part. This seems to be also the most natural for a real-word engine since the two threads use the same core engine (e.g., same term weighting function).

\end{abstract}



\received{Feb. 2023}

\maketitle

\section{Introduction}



In information retrieval, system effectiveness can be optimized by adjusting many parameters and components the system consists of. For example, participants to shared tasks such as TREC~\footnote{Text REtrivial Conference, https://trec.nist.gov} generally consider a system where the parameters of the term weighting function and the ones of the query expansion model are optimized for a given collection~\cite{song1999general,ma2022scattered}. Similarly, in real-world system, the component parameters can be optimized based on past queries or users' implicit feedback~\cite{ai2018unbiased,gao2019fair}. 

Alternatively, the query processing thread can be adapted to each query. For example, selective query expansion (SQE) uses one of two alternative query processing threads: the original query or the automatically expanded query~\cite{amati2004query,cronen2004framework,he2007combining,zhao2012automatic}. 
Selective query processing (SQP) generalizes SQE to various query processing threads. Here, the threads may use different term weighting schemes or different query expansion models~\cite{he2004query,Xu2016,deveaud2018learning}. SQP thus differs from selective search which aims to improve the response time by limiting the part of the data the system is searching from, like in shard selection, for example~\cite{baeza2009efficiency,aly2013taily}. In SQP, the optimization takes place in the processing threads and retrieval component parameters.

For SQE, the decision on which thread should process the query is based on some query features. Cronen-Townsend \textit{et al.}~\cite{cronen2004framework} and Amati \textit{et al.}~\cite{amati2004query} estimated how much the results of the expanded query strayed away from the sense of the original one to make the decision. Xu \textit{et al.}~\cite{xu2009query} improved SQE such that the system chooses between more options and considered different query expansion strategies according to the query type. They defined three types of queries: ambiguous queries, queries about a specific entity, and broader queries, and proposed different methods to treat each type. Queries are categorized in an unsupervised way, based on Wikipedia page types. Kevyn Collins-Thompson~\cite{collins2009reducing} introduced a constraint optimization framework for SQE by reducing the risk of query expansion. He et al. clustered the queries according to query features such as their length or term ambiguity and then associated the best processing thread to each cluster through a training phase~\cite{he2004query}. Deveaud et al. learned the best processing thread using learning to rank principle, where threads are ranked for each query according to their effectiveness~\cite{deveaud2018learning}. Mothe et al. \cite{mothe2021defining} used a risk-based approach to decide which threads should be included in the meta-system. The risk-reward trade-off function they defined adds a configuration if it increases system effectiveness on some queries and at the same time minimizes the risk of poor performance of that configuration if it is chosen for other queries.

SQE/SQP mainly focuses on effectiveness but seldom mentions efficiency. We believe that both are important criteria to consider. This paper thus examines selective query processing in different settings, both in terms of effectiveness and efficiency. We conduce intensive experiments using three TREC adhoc collections (TREC78, GOV2, and WT10g) and three effectiveness measures (AP, nDCG@10, and P@10). We compare the results to different baselines, including fusion techniques and best trained. In our experiments, candidate query processing threads vary in the weighting function used, the query expansion model as well as the number of documents and terms used in the query expansion. Candidate query processing threads are defined based on a risk-based function~\cite{mothe2021defining}. The selective model is trained using a supervised approach that considers query features. For efficiency, similar to Kulkarni and Callan~\cite{kulkarni2015selective}, we measure the elapsed time for each processing steps: data generation ({Gen}), model training ({Train}), and runtime processing ({RT}).

The contributions of this paper are three folds:
\begin{itemize}
    \item First, we analyze the selective query processing methods in terms of both effectiveness and efficiency on three adhoc reference collections when two query processing threads are used in different settings;
    
    \item Second, we compare the performance to different baselines (BM25, learning to rank, best trained, and CombSum) and show that selective query processing outperforms all our baselines both in terms of effectiveness and efficiency;
    \item Third, when comparing selective query processing methods, we found that the best trade-off between effectiveness and efficiency is obtained when using the best trained processing thread and its query expansion counterpart. This seems to be also the most natural for a real-word engine since the two threads use the same core engine (e.g., same term weighting function).
\end{itemize}


\section{Methodology}
\label{sec:methodology}

\subsection{Query processing threads and grid of points}
Processing threads can differ according to the term weighting schema, the type of expansion model used, the number of documents and expansion terms used. A query processing thread is a set ($W$,$Q$,$D$,$T$), where $W$ is the term weighting function, $Q$ is the query expansion model, $D$ is the number of documents for query expansion, and $T$ is the number of added terms in query expansion. For $W$ and $Q$, we consider all the models implemented in  Terrier\footnote{\url{http://terrier.org/docs/current/javadoc/org/terrier/matching/models/package-summary.html}}. $D$ and $T$ can vary from 0 (no expansion) to a defined number. We build all the possible comprehensive query processing threads using the values as detailed in Table~\ref{tab:paravalues}. This results in more than $20,000$ query processing threads. We used all the query processing threads to perform batch retrieval on the topics. The parameters of the query processing threads and the estimated effectiveness measures produce a grid of points~\cite{ferro2016clef}.

\begin{table}[!htbp]
\caption{Parameters of the query processing threads}{
\begin{tabular}{@{}l@{}}
\textbf{Weighting model $W$}\\
\hline
BB2, BM25, DFRee, DirichletLM, HiemstraLM,InB2, InL2, JsKLs,\\
PL2, DFI0, XSqrAM, DLH13, DLH, DPH, IFB2, TFIDF, InexpB2, \\
DFRBM25, LGD, LemurTFIDF, InexpC2\\[3pt]
\textbf{Query expansion model $Q$}\\
\hline
None, KL, Bo1, Bo2, KLCorrect, Information, KLComplete\\[3pt]
\textbf{QE Hyperparameter and Values}\\
\hline
\# of Expansion document $D$: 0, 2, 5, 10, 20, 50, 100\\
\# of Expansion terms $T$: 0, 2, 5, 10, 15, 20
\end{tabular}}
\label{tab:paravalues}
\end{table}

\subsection{Candidate threads}
\label{Sec:Candidate_threads}
To select the candidate threads from the grid of points, we use the $E_{Risk}$ function as defined by Mothe and Ullah~\cite{mothe2021defining}. Their function aims at selecting $k$ candidate query processing threads. We re-implemented it so that it can fit our purpose of selecting two threads, either for SQE or SQP. 
The first candidate $C_1$: ($W_1$, $Q_1$, $D_1$, $T_1$) is the thread that gets the highest effectiveness -considering the chosen measure- in average over the training query set. It is used in both SQE ans SQP. 
The second candidate $C_2$: ($W_2$, $Q_2$, $D_2$, $T_2$) may differ for SQE and SQP.

For SQE, if $Q_1$ $=$ ``None", then the second candidate will correspond to the same thread with query expansion. That is, $C_2$: ($W_1$, $Q_2$, $D_2$, $T_2$). That is to say, the same weighting model $W_1$, but the next best thread with query expansion will be selected. 
Reversely, if the best thread includes query expansion ($Q_1$ $\neq$ ``None"), the second query processing thread will not use query expansion $C_2$: ($W_1$, None, 0, 0).

For SQP, given the first thread and grid of points, we select the second thread by maximizing the overall gain for retrieval effectiveness, following the $E_{Risk}$ function~\cite{mothe2021defining}. We estimate risk, reward, and gain for threads in grid of points and choose the one which obtains the maximum gain. It is thus a greedy diversification technique that optimizes effectiveness.

\subsection{Candidate selection}
Following in Deveaud \textit{et al.}~\cite{deveaud2018learning} or in Mothe and Ullah~\cite{mothe2021defining}, the query-thread association problem is cast as a ranking problem where L2R algorithms (e.g., SVM-rank) are applied to decide which of the two threads should be used by the meta-system (See~\ref{sub:training}).

\section{Experiments}
\label{sec:Experiments}

\subsection{\textbf{Datasets}}
We report here results on three 
standard TREC collections, TREC78, GOV2, and WT10g from the adhoc task. 
For TREC78, there are approximately 500K newspaper articles. GOV2 collection includes 25 million web pages from .gov domain. The WT10G collection is composed of approximately 1.6 million Web/Blog page documents. The TREC test collections include a set of topics. The ``standard" format of a TREC topic statement comprises a topic ID, a title, a description, and a narrative. We consider the topic title that contains two or three words, on average as the query to process. 
There are 100 topics in TREC78 (merging topics from TREC7 and TREC8), 150 in GOV2, and 100 in WT10g. 
The collections provide \textit{qrels} (i.e., judged documents) 
which are used in the evaluation program
\textit{trec\_eval}\footnote{http://trec.nist.gov/trec\_eval/} to calculate the effectiveness measures for adhoc metrics.

\subsection{\textbf{Evaluation protocol}}
\label{sub:eval-protocol}
We used two-fold cross-validation to ensure independent sets and keep enough samples in both the training and test sets~\cite{wong2017dependency}. We proceeded as follows: half of the queries, let us call this query set $Q_A$, are used for training, while the other half, $Q_{\overline A}$, is used for testing. Then, reversely, $Q_{\overline A}$ is used for training, while $Q_A$ is used for testing. We made three random draws to split the query set 
into $Q_A$ and $Q_{\overline A}$. %
To prepare the folds for three trials, we shuffled the query set using the R's sample function (random seed=42) and divided the query set into two subsets for each trial. We report the average and standard deviation across three trials and the two test folds. Thus, each mean and standard deviation is over a total of six measurements. The same splits are used in whatever method needs training. We use a two-tailed paired t-test with Bonferroni correction (p-value $<$ 0.05) for statistical significance analysis.


\subsection{\textbf{Query processing thread selection.}} To compare the combination methods, we consider the three settings (the two first have been described in detail in Section~\ref{Sec:Candidate_threads}) 
 as follows:
\begin{itemize}
    \item In {E$_{Risk}$-SQE} we take the first query processing thread selected by E$_{Risk}$ function and its query expansion/non-expansion counterpart. 
   \item  In {E$_{Risk}$-SQP} we take the first two search threads selected by E$_{Risk}$ function. The chosen threads may or may not have a query expansion component.
   \item In {Best-SQE}, we take the best thread overall and its query expansion/non-expansion counterpart. 
In each case, we have two query processing threads that are then selected by SQE based on SVM-rank machine learning.
\end{itemize}

\subsection{\textbf{Features and labels for model training}}
\label{sub:training}
Query features are required to train some models, specifically for L2R-document and selective approaches. We opted for LETOR features that were used for document ranking models~\cite{qin2010letor}. To extract the LETOR features, we use the FAT component of Terrier IR~\cite{macdonald2013learning}. The FAT component stores the posting lists of all the documents that make the top-K retrieved set in memory. Different features are computed using these posting lists. We calculate LETOR features directly from BM25 and aggregate the query-document features as Chifu et al. successfully did for query performance prediction~\cite{chifu2018query}. 
Then, the query vector is labeled by the effectiveness of the query processing thread applied to that query.

\subsection{\textbf{Evaluation metrics}}
We consider the effectiveness and efficiency-based evaluation metrics to compare the combination methods.
\textit{Effectiveness metrics}: 
We considered the Precision at the cut-off 10 documents (P@10), Average Precision (AP)~\cite{carterette2011overview,sakai2007reliability}, normalized Discounted Cumulative Gain at the cut-off 10 documents (nDCG@10)~\cite{jarvelin2017ir}. \\
\textit{Efficiency metrics}:  
Running a query processing thread requires several processing steps, such as data pre-processing, batch retrieval, model training, prediction, and eventually re-ranking. To estimate the efficiency of a process, we measure the elapsed time (T$_{Total}$) when accomplishing each underlying process. 

    $T_{Total}$: a total of elapsed time at accomplishing a process.

Kulkarni and Callan~\cite{kulkarni2015selective} used a similar approach to estimate the efficiency in selective search. We categorize the different processes into three groups: data generation ({Gen}), model training ({Train}), and main processing (Run-time (RT)). The {Gen} and {Train} groups are offline processes and the main processing is a run-time process. In the {Gen} group, the usual process involves query processing thread selection, batch retrieval, feature extraction, etc. In the {Train} group, the process involves training the machine learning model. The main processing includes feature extraction, models' prediction, retrieval process using different query processing threads, result fusion (Data fusion), per-query retrieval, etc. The combination methods we evaluate do not necessarily require all these processes.
 
\subsection{\textbf{Baselines and compared methods}}
\label{subsec:baseline}
We consider the {BM25}, {L2R-D SVM$^r$}, {Best trained}, and CombSum as baselines. 
\begin{itemize}
    \item {BM25} is obtained using Terrier BM25 with default parameters b=0.75 and k=1.2 (default hyperparameter values)~\cite{robertson2009probabilistic} on the entire set of the queries;
    \item {L2R-D SVM$^r$} is the usual learning to rank (L2R) documents where the initial ranking is obtained using BM25. The L2R is learned based on the SVM-rank. 
    \item {Best trained} is the best query processing thread among the grid of points, the one that maximized the effectiveness measure considered: it is chosen for $Q_{A}$ and applied to $Q_{\overline{B}}$, and vice versa.
    \item {CombSum}: It implements data fusion as the combinations of system document lists; we consider the CombSum~\cite{fox1994combination} on the two configurations selected by $E_{Risk}$-SQE.
\end{itemize}

\section{Results and discussion}
\label{sec:Results}

We evaluate the results on baselines, E$_{Risk}$-SQE, E$_{Risk}$-SQP, and Best-SQE (See Tables~\ref{tab:trec78_sqe},~\ref{tab:gov2_sqe}, and~\ref{tab:wt10g_sqe}) for TREC78, GOV2, and WT10g, respectively. The first four rows refer to the baseline models, BM25, L2R-Document (L2R-D SVM$^{r}$), the Best trained, and CombSum. Then, the next rows denote the E$_{Risk}$-SQE, the SQP for E$_{Risk}$-SQP, and the Best-SQE. 
The SQP ranking function is learned using SVM-rank technique.
The left-side of the tables shows the effectiveness metrics (AP, nDCG@10, and P@10) while the right-side reports the efficiency metrics in terms of elapsed time when accomplishing a process (Gen, Train, and Run-time) in milliseconds for each query in average.

\begin{table*}[!]
\caption{Effectiveness and efficiency for SQP on TREC78. $\vartriangle$ and ${\uparrow}$ indicate statistically significant improvement of SQP over baselines L2R documents and Best trained - paired t-test ($p < 0.05$). Training was based on SVM-rank algorithm. Effectiveness is on the left side and efficiency on the right side for Generation (Gen), Training (Train), and Run-time phases. The mean and [standard deviation] for 3 trials are reported.}
\vspace{-3mm}
\begin{tabular}{@{}l@{\hspace{.1em}}l@{} c@{\hspace{.5em}} c@{\hspace{.5em}} c|c  r@{\hspace{.5em}} c@{\hspace{.5em}} r@{}}
& & \multicolumn{3}{c}{\textbf{Effectiveness}} && \multicolumn{3}{c}{\textbf{Efficiency (ms)}}\\
\hline
& Methods & \textbf{AP} & \textbf{nDCG@10} & \textbf{P@10} && \textbf{Gen} & \textbf{Train} & \textbf{RT}\\
\hline
\multicolumn{4}{@{}l}{\textbf{Baselines}}\\
& BM25 & .211$^{~~~~~~~~~~}$ & .465$^{~~~~~~~~~~}$ & .448$^{~~~~~~~~~~}$ && - & - & 40.49\\
& L2R-D SVM$^r$ & .216 [.000]$^{~~}$ & .478 [.001]$^{~~}$ & .459 [.004]$^{~~}$ & & 3601 [9.5] & 25.1 [17] & 673 [14]\\
& Best trained & .240 [.003]$^{~~}$ & .525 [.004]$^{~~}$ & .486 [.007]$^{~~}$ & & - & 367 [5.8] & 425 [23]\\
& CombSUM & \textbf{.256}$^{\vartriangle\uparrow}$ [.003] & .496 [.006]& .467 [.010] && 1450 [3] & - & 1056 [50]\\
\hline
\multicolumn{4}{@{}l}{\textbf{Selective query expansion and processing}}\\
& E$_{Risk}$-SQE & .248$^{\vartriangle\uparrow}$ [.002] & .503 [.027] & \textbf{.503}$^{\vartriangle\uparrow}$ [.003] && 1535 [42] & .93 [.23] & \textbf{45} [1.5]\\
& E$_{Risk}$-SQP & .247$^{\vartriangle\uparrow}$ [.002] & \textbf{.530}$^{\vartriangle}$ [.002] & {.500}$^{\vartriangle}$ [.005] && 1091 [27] & .67 [.11] & \textbf{46} [1.7]\\
& Best-SQE &  .248$^{\vartriangle\uparrow}$ [.002] & .499 [.028] & \textbf{.503}$^{\vartriangle\uparrow}$ [.003] && 1049 [23] & .78 [.08] & \textbf{46} [2.1]
\end{tabular}\label{tab:trec78_sqe}
\end{table*}

\begin{table*}[!]
\caption{Effectiveness and efficiency for  SQP on GOV2. The notations are the same as in Table~\ref{tab:trec78_sqe}.}
\vspace{-3mm}
\begin{tabular}{@{}l@{\hspace{.1em}}l@{} c@{\hspace{.5em}} c@{\hspace{.5em}} c|@{}c  r@{\hspace{.5em}} c@{\hspace{.5em}} r@{}}
& & \multicolumn{3}{c}{\textbf{Effectiveness}} && \multicolumn{3}{c}{\textbf{Efficiency (ms)}}\\
\hline
& Methods & \textbf{AP} & \textbf{nDCG@10} & \textbf{P@10} && \textbf{Gen} & \textbf{Train} & \textbf{RT}\\
\hline
\multicolumn{4}{@{}l}{\textbf{Baselines}}\\
& BM25 & .269$^{~~~~~~~~~~}$ & .458$^{~~~~~~~~~~}$ & .527$^{~~~~~~~~~~}$ && - & - & 294.67\\
& L2R-D SVM$^r$ & .279 [.000]$^{~~}$ & .491 [.002]$^{~~}$ & .567 [.003]$^{~~}$ && 75842 [21] & 20 [.24] & 625 [4.5]\\
& Best trained & .351 [.005]$^{~~}$ &.487 [.012]$^{~~}$ & .590 [.010]$^{~~}$ && - & 364 [1.9] & 9556 [662]\\
& CombSUM & .358$^{\vartriangle}$ [.003]& .553$^{\vartriangle\uparrow}$ [.003]& .645$^{\vartriangle\uparrow}$ [.001] && 1441 [.9] & & 14700 [670]\\
\hline
\multicolumn{4}{@{}l}{\textbf{Selective query expansion and processing}}\\
& E$_{Risk}$-SQE & \textbf{.362}$^{\vartriangle\uparrow}$ [.001] & .540$^{\vartriangle\uparrow}$ [.004] & \textbf{.649}$^{\vartriangle\uparrow}$ [.004] && 1540 [50] & .55 [.07] & 3190 [1273]\\
& E$_{Risk}$-SQP & \textbf{.363}$^{\vartriangle\uparrow}$ [.002] & .538$^{\vartriangle\uparrow}$ [.006] & \textbf{.649}$^{\vartriangle\uparrow}$ [.003] && 1119 [46] & .65 [.16] & \textbf{2121} [24]\\
& Best-SQE & \textbf{.362}$^{\vartriangle\uparrow}$[.001] & .540$^{\vartriangle\uparrow}$[.004] & \textbf{.649}$^{\vartriangle\uparrow}$[.004] && 968 [15] & .62 [.14] & \textbf{2090} [34]\\
\end{tabular}\label{tab:gov2_sqe}
\end{table*}

\begin{table*}[!]
\caption{Effectiveness and efficiency for SQP on WT10g. The notations are the same as in Table~\ref{tab:trec78_sqe}.}
\vspace{-3mm}
\begin{tabular}{@{}l@{\hspace{.1em}}l@{} c@{\hspace{.5em}} c@{\hspace{.5em}} c|@{}c  r@{\hspace{.5em}} c@{\hspace{.5em}} r@{}}
& & \multicolumn{3}{c}{\textbf{Effectiveness}} && \multicolumn{3}{c}{\textbf{Efficiency (ms)}}\\
\hline
& Methods & \textbf{AP} & \textbf{nDCG@10} & \textbf{P@10} && \textbf{Gen} & \textbf{Train} & \textbf{RT}\\
\hline
\multicolumn{4}{@{}l}{\textbf{Baselines}}\\
& BM25 & .172$^{~~}$$^{~~}$ & .288$^{~~}$$^{~~}$ & .273$^{~~}$$^{~~}$ && - & - & 249.77\\
& L2R-D SVM & .192 [.002]$^{~~}$ & .330 [.002]$^{~~}$ & .310 [.005]$^{~~}$ && 7144 [5] & 18 [.69] & 690 [1.0]\\
& Best trained & .224 [.006]$^{~~}$ & .401 [.002]$^{~~}$ & .396 [.012]$^{~~}$ && - & 345 [23] & 1842 [278]\\
& CombSUM & .267$^{\vartriangle}$$^{\uparrow}$ [.004] & .437$^{\vartriangle}$$^{\uparrow}$ [.010] & .418$^{\vartriangle}$$^{\uparrow}$ [.006] && 1372 [23] & & 3736 [270]\\
\hline
\multicolumn{4}{@{}l}{\textbf{Selective query expansion and processing}}\\
& E$_{Risk}$-SQE & .242$^{\vartriangle}$$^{\uparrow}$$^{~~}$ [.014]& .388$^{\vartriangle}$$^{~~}$ [.028] & .376$^{\vartriangle}$$^{~~}$ [.015] && 1574 [55] & .83 [.12] & 722 [244]\\
& E$_{Risk}$-SQP & \textbf{.255}$^{\vartriangle}$$^{\uparrow}$$^{~~}$ [.009] & \textbf{.414}$^{\vartriangle}$$^{~~}$$^{~~}$ [.010] & \textbf{.395}$^{\vartriangle}$$^{~~}$$^{~~}$ [.005] && 1085 [19] & .73 [.09] & \textbf{329} [44]\\
& Best-SQE & .242$^{\vartriangle}$$^{\uparrow}$$^{~~}$ [.014] & .400$^{\vartriangle}$$^{~~}$ [.034] & .374$^{\vartriangle}$$^{~~}$ [.025] && 1021 [43] & .69 [.08] & \textbf{325} [135]\\
\end{tabular}\label{tab:wt10g_sqe}
\end{table*}

First, we can observe that BM25 is consistently less effective than the other non-combined approaches (See first row, Tables~\ref{tab:trec78_sqe} to \ref{tab:wt10g_sqe}). Although this result was expected, it is worth to check that simpler methods do not achieve better results. Among baselines, the CombSum is the only one that does not need any training (4th row). It achieves, however, better results than the baselines that need training on GOV2 and WT10g collections in terms of effectiveness. The CombSum is not efficient because it implies to run each query twice and then to fuse the results. Among baselines, best-trained offers a good trade-off. 

Considering the selective query processing approaches, they are consistently more effective than non-trained baselines across measures and collections (statistically significant). Comparing the three settings, the SQP approach is slightly more effective than the SQE approach on WT10G, however Best-SQE is slightly more effective on the two other collections. Best-SQE is the most efficient at run time; thus could be considered as the best option with regard to effectiveness vs. efficiency trade-off. Meaning that the query expansion part of a thread should be optimized rather than the other system components such as the weighting function. This result somehow contradicts previous results that show that the weighting model is a key factor~\cite{ayter2015statistical,ferro2018toward}. However, here a training step is used to select the best processing thread that should be effective for many queries. We are not considering the best component parameters for each query individually, but rather the two best processing threads for a set of queries.

All SQE or SQP methods require training time. From the result, we can observe that the SQP with SVM-rank ($E_{Risk}-SQP$) achieves significantly better performance over the baselines in terms of both effectiveness and efficiency metrics. 
These findings are consistent in both Tables~\ref{tab:trec78_sqe} and~\ref{tab:gov2_sqe}. The reason is that most of the time, a second processing thread that does not use query expansion is more effective than a query expansion one. 


\section{Conclusions}
In this paper we aimed to compare selective query processing (SQP) settings considering different threads. We found that all settings are more effective than the baselines.
Our results contradict the intuition that data fusion was more effective although less efficient (CombSum baseline). 
We found that selective approach is both more effective and efficient than data fusion.

With regard to the different selective query processing settings, the best effectiveness vs. efficiency trade-off goes to the selective query processing (SQP) where one thread is selected as the best over the set of queries and the second as a query expansion counterpart.

Here we consider two query processing threads. But both data fusion and selective query processing could handle more than two. In our future work, we would like to investigate the combination of more than two query processing threads.
\balance
\bibliographystyle{ACM-Reference-Format}
\bibliography{References}
\end{document}